\documentclass[12pt]{article}
 
\usepackage{amsmath,amssymb}
\usepackage{bm}
\usepackage{graphicx, color}
\usepackage{wrapfig}
\usepackage{cite}
\begin{document}
\title{
\begin{flushright}
\ \\*[-80pt] 
\begin{minipage}{0.22\linewidth}
\normalsize
APCTP Pre2019-011
 \\*[50pt]
\end{minipage}
\end{flushright}
{\Large \bf 
 Towards  unification of  quark and lepton flavors \\
 in  $A_4$ modular  invariance
\\*[20pt]}}

\author{ 
\centerline{
Hiroshi Okada $^{a,b}\footnote{E-mail address: hiroshi.okada@apctp.org}$~ and 
~~Morimitsu Tanimoto $^{c}\footnote{E-mail address: tanimoto@muse.sc.niigata-u.ac.jp}$} \\*[5pt]
\centerline{
\begin{minipage}{\linewidth}
\begin{center}
$^a${\it \normalsize
Asia Pacific Center for Theoretical Physics, Pohang 37673, Republic of Korea} \\*[5pt]
$^b${\it \normalsize
Department of Physics, Pohang University of Science and Technology, Pohang 37673,\\ Republic of Korea} \\*[5pt]
$^c${\it \normalsize
Department of Physics, Niigata University, Niigata 950-2181, Japan}
\end{center}
\end{minipage}}
\\*[50pt]}

\date{
\centerline{\small \bf Abstract}
\begin{minipage}{0.9\linewidth}
\medskip 
\medskip 
\small 
We study quark and lepton mass matrices in  the  $A_4$ modular symmetry towards the unification of the quark and lepton flavors.
 We adopt modular forms of weights $2$ and  $6$
 for  quarks and charged leptons, while
 we use modular forms of weight $4$ for the neutrino mass matrix
 which is  generated by the Weinberg operator.
 We obtain the successful quark mass matrices, in which
 the down-type quark mass matrix is constructed by
 modular forms of weight $2$, but  the up-type quark mass matrix is constructed by modular forms of weight $6$.
  The viable region of  $\tau$  is close to $\tau=i$.
  Lepton mass matrices also work well  at nearby  $\tau=i$, which 
 overlaps with the one of  the quark sector, for the normal hierarchy of neutrino masses. 
 In the common $\tau$ region for quarks and leptons,
 the predicted  sum of neutrino masses is  $87$--$120$\,meV
  taking account of  its  cosmological bound.
 Since both the Dirac CP phase $\delta_{CP}^\ell$ and $\sin^2\theta_{23}$
 are correlated with the sum of neutrino masses,
  improving its cosmological bound 
 provides  crucial tests for our scheme
 as well as  the precise measurement of $\sin^2\theta_{23}$ and 
 $\delta_{CP}^\ell$.
 The effective neutrino mass of the $0\nu\beta\beta$ decay 
 is   $\langle m_{ee}\rangle=15$--$31$\,meV. 
  It is remarked that  the modulus $\tau$ is fixed at nearby $\tau=i$
   in the fundamental domain of SL$(2,Z)$,
  which suggests  the  residual symmetry $Z_2$ in the quark and lepton mass matrices.
 The inverted  hierarchy of  neutrino masses is excluded
  by the cosmological bound of the sum of neutrino masses.
\end{minipage}
}

\begin{titlepage}
\maketitle
\thispagestyle{empty}
\end{titlepage}

\section{Introduction}

The standard model (SM) was well established by the discovery of
 the Higgs boson.
However, the flavor theory  of quarks and leptons is still unknown.
In order to understand the origin of the flavor structure,
many works have appeared by using the  discrete  groups
 for flavors.
 In  the early models of quark masses and mixing angles, 
the  $S_3$ symmetry was used 
 \cite{Pakvasa:1977in,Wilczek:1977uh}.
 It was also discussed  to understand the large mixing angle
 \cite{Fukugita:1998vn} in the oscillation of atmospheric neutrinos \cite{Fukuda:1998mi}. 
 For the last twenty years, the  discrete symmetries of flavors have been developed, that is
  motivated by the precise observation of  flavor mixing angles of  leptons
  \cite{Altarelli:2010gt,Ishimori:2010au,Ishimori:2012zz,Hernandez:2012ra,King:2013eh,King:2014nza,Tanimoto:2015nfa,King:2017guk,Petcov:2017ggy}.

Many models have been proposed by using 
the non-Abelian discrete groups  $S_3$, $A_4$, $S_4$, $A_5$ and other groups with larger orders to explain the large neutrino mixing angles.
Among them, the $A_4$ flavor model is an attractive one 
because the $A_4$ group is the minimal one including a triplet 
 irreducible representation, 
which allows for a natural explanation of the  
existence of  three families of leptons 
\cite{Ma:2001dn,Babu:2002dz,Altarelli:2005yp,Altarelli:2005yx,
Shimizu:2011xg,Petcov:2018snn,Kang:2018txu}.
However, the variety of models is so wide that it is difficult to obtain 
a clear evidence of the $A_4$ flavor symmetry.

Recently, a new  approach to the lepton flavor problem appeared
based on the invariance under the modular group \cite{Feruglio:2017spp}, 
where the model of the finite
modular group  $\Gamma_3 \simeq A_4$ has been presented.
This work  inspired further studies of the modular invariance approach
 to the lepton flavor problem. 
It should be emphasized that there is a significant difference between the 
models based on the $A_4$ modular symmetry and those based on the usual 
non-Abelian discrete $A_4$ flavor symmetry.
Yukawa couplings transform non-trivially under the modular group
and are written in terms of modular forms which are  
holomorphic functions of   the modulus  $\tau$.

The modular group includes the finite groups $S_3$, $A_4$, $S_4$, and $A_5$ \cite{deAdelhartToorop:2011re}.
Therefore, an interesting framework for the construction of flavor models
has been put forward based on  the $\Gamma_3 \simeq A_4$ modular group \cite{Feruglio:2017spp}, and further, based on $\Gamma_2 \simeq S_3$ \cite{Kobayashi:2018vbk}.
The proposed flavor models with modular symmetries  
$\Gamma_4 \simeq S_4$ \cite{Penedo:2018nmg} 
and  $\Gamma_5 \simeq A_5$ \cite{Novichkov:2018nkm} 
 have also  stimulated  studies of  flavor structures of quarks and leptons.
Phenomenological discussions of the neutrino flavor mixing have been done
based on  $A_4$ \cite{Criado:2018thu,Kobayashi:2018scp}, $S_4$ \cite{Novichkov:2018ovf,Kobayashi:2019mna}, $A_5$ \cite{Ding:2019xna}, 
and  $T'$~\cite{Liu:2019khw} modular groups, respectively.
In particular, the comprehensive analysis of the $A_4$ modular group 
has provided a distinct prediction of the neutrino mixing angles and the CP violating phase \cite{Kobayashi:2018scp}.
The $A_4$ modular symmetry has been also applied to the $SU(5)$ grand
unified theory (GUT) of  quarks and leptons  \cite{deAnda:2018ecu},
while the residual symmetry of the $A_4$ modular symmetry has been investigated phenomenologically \cite{Novichkov:2018yse}.
Furthermore, modular forms for $\Delta(96)$ and $\Delta(384)$ were constructed \cite{Kobayashi:2018bff},
and the extension of the traditional flavor group  is discussed with modular symmetries \cite{Baur:2019kwi}.
Moreover, multiple modular symmetries are proposed as the origin of flavor\cite{deMedeirosVarzielas:2019cyj}.
The modular invariance has been also studied combining with the generalized CP symmetries for theories of flavors \cite{Novichkov:2019sqv}.
 The quark mass matrix  has been discussed in the $S_3$ and $A_4$ modular symmetries as well\cite{Kobayashi:2018wkl,Kobayashi:2019rzp,Okada:2018yrn}.
 Besides mass matrices of quarks and leptons,
 related topics have been discussed 
  in the baryon number violation  \cite{Kobayashi:2018wkl}, 
   the dark matter \cite{Nomura:2019jxj, Okada:2019xqk}
   and the modular symmetry anomaly  \cite{Kariyazono:2019ehj}.
 
 
In this work, we study both quark and lepton mass matrices
in the $A_4$ modular symmetry.
 If flavors of quarks and leptons are originated from a same two-dimensional compact space, 
  quarks and leptons have the same flavor symmetry and the same value of 
 the modulus $\tau$.
  Therefore, it is challenging to reproduce
observed three Cabibbo-Kobayashi-Maskawa (CKM) mixing angles and the CP violating phase 
while observed large mixing angles  in the lepton sector
within the framework of the $A_4$ modular invariance with the common $\tau$.
This work provides a new aspect in order to discover the unification theory
of the quark and lepton flavors.
We have already discussed the quark mass matrices in terms of $A_4$
 modular forms of weight $2$ \cite{Okada:2018yrn}.  It has been found that 
  quark mass matrices of $A_4$ do not work
  unless Higgs sector is extended by  $A_4$ triplet representations.
 In this paper, we propose to adopt modular forms of weight $6$
 in addition to  modular ones of weight $2$ for quarks
 with  the Higgs sector of SM.
  We  use  modular forms of weight $4$ for the neutrino mass matrix,
  which is generated by the Weinberg operator.
 The common value of the modulus $\tau$ is  successfully obtained
  by using observed four CKM matrix elements and three lepton mixing angles.  
 We also predict the CP violating Dirac phase  of  leptons,
 which is expected to be observed at T2K and NO$\nu$A experiments \cite{T2K:2020,Adamson:2017gxd}, and the sum of neutrino masses
 \footnote{In order to reduce the number of  parameters, 
 	we   discussed another $A_4$ quark/lepton mass matrices
 	after this work \cite{Okada:2020rjb}.
 	However, the sum of neutrino masses is predicted to be larger than
 	the cosmological upper-bound, $120$\,meV.
 	Indeed, a model with two parameters fewer than the present model gives 
 	$140$\,meV for  the sum of neutrino masses.}.


The paper is organized as follows.
In section 2,  we give a brief review on the modular symmetry and 
modular forms of weights $2$, $4$ and $6$. 
In section 3, we present the model for quark mass matrices in the $A_4$
modular symmetry.
In section 4, we show  numerical results for the CKM matrix.
In section 5, we discuss the lepton mass matrices and present numerical results. 
Section 6 is devoted to a summary.
In Appendix A, the tensor product  of the $A_4$ group is presented.
In Appendix B, we present how to obtain  Dirac $CP$ phase, Majorana phases and  effective mass of the $0\nu\beta\beta$ decay.

\section{Modular group and modular forms of weights $2$, $4$, $6$}

The modular group $\bar\Gamma$ is the group of linear fractional transformations
$\gamma$ acting on the modulus  $\tau$, 
belonging to the upper-half complex plane as:
\begin{equation}\label{eq:tau-SL2Z}
\tau \longrightarrow \gamma\tau= \frac{a\tau + b}{c \tau + d}\ ,~~
{\rm where}~~ a,b,c,d \in \mathbb{Z}~~ {\rm and }~~ ad-bc=1, 
~~ {\rm Im} [\tau]>0 ~ ,
\end{equation}
 which is isomorphic to  $PSL(2,\mathbb{Z})=SL(2,\mathbb{Z})/\{I,-I\}$ transformation.
This modular transformation is generated by $S$ and $T$, 
\begin{eqnarray}
S:\tau \longrightarrow -\frac{1}{\tau}\ , \qquad\qquad
T:\tau \longrightarrow \tau + 1\ ,
\label{symmetry}
\end{eqnarray}
which satisfy the following algebraic relations, 
\begin{equation}
S^2 =\mathbb{I}\ , \qquad (ST)^3 =\mathbb{I}\ .
\end{equation}

 We introduce the series of groups $\Gamma(N)~ (N=1,2,3,\dots)$,
   called principal congruence subgroups, defined by
 \begin{align}
 \begin{aligned}
 \Gamma(N)= \left \{ 
 \begin{pmatrix}
 a & b  \\
 c & d  
 \end{pmatrix} \in SL(2,\mathbb{Z})~ ,
 ~~
 \begin{pmatrix}
  a & b  \\
 c & d  
 \end{pmatrix} =
  \begin{pmatrix}
  1 & 0  \\
  0 & 1  
  \end{pmatrix} ~~({\rm mod} N) \right \}
 \end{aligned} .
 \end{align}
 For $N=2$, we define $\bar\Gamma(2)\equiv \Gamma(2)/\{I,-I\}$.
Since the element $-I$ does not belong to $\Gamma(N)$
  for $N>2$, we have $\bar\Gamma(N)= \Gamma(N)$.
   The quotient groups defined as
   $\Gamma_N\equiv \bar \Gamma/\bar \Gamma(N)$
  are  finite modular groups.
In this finite groups $\Gamma_N$, $T^N=\mathbb{I}$  is imposed.
 The  groups $\Gamma_N$ with $N=2,3,4,5$ are isomorphic to
$S_3$, $A_4$, $S_4$ and $A_5$, respectively \cite{deAdelhartToorop:2011re}.

Modular forms of  level $N$ are 
holomorphic functions $f(\tau)$  transforming under 
$\Gamma(N)$ as:
\begin{equation}
f(\gamma\tau)= (c\tau+d)^kf(\tau)~, ~~ \gamma \in \Gamma(N)~ ,
\end{equation}
where $k$ is the so-called as the  modular weight.

Superstring theory on the torus $T^2$ or orbifold $T^2/Z_N$ has the modular symmetry \cite{Lauer:1989ax,Lerche:1989cs,Ferrara:1989qb,Cremades:2004wa,Kobayashi:2017dyu,Kobayashi:2018rad}.
Its low energy effective field theory is described in terms of  supergravity theory,
and  string-derived supergravity theory has also the modular symmetry.
Under the modular transformation of Eq.(\ref{eq:tau-SL2Z}), chiral superfields $\phi^{(I)}$ 
transform as \cite{Ferrara:1989bc},
\begin{equation}
\phi^{(I)}\to(c\tau+d)^{-k_I}\rho^{(I)}(\gamma)\phi^{(I)},
\end{equation}
where  $-k_I$ is the modular weight and $\rho^{(I)}(\gamma)$ denotes a unitary representation matrix of $\gamma\in \bar\Gamma$.

 In the present article we study global supersymmetric models, e.g., 
minimal supersymmetric extensions of the Standard Model (MSSM).
The superpotential which is built from matter fields and modular forms
is assumed to be modular invariant, i.e., to have 
a vanishing modular weight. For given modular forms 
this can be achieved by assigning appropriate
weights to the matter superfields.

The kinetic terms  are  derived from a K\"ahler potential.
The K\"ahler potential of chiral matter fields $\phi^{(I)}$ with the modular weight $-k_I$ is given simply  by 
\begin{equation}
K^{\rm matter} = \frac{1}{[i(\bar\tau - \tau)]^{k_I}} |\phi^{(I)}|^2,
\end{equation}
where the superfield and its scalar component are denoted by the same letter, and  $\bar\tau =\tau^*$ after taking the vacuum expectation value (VEV)
\footnote{The most general K\"ahler potential consistent with the modular symmetry  possibly contains additional terms, as
	recently pointed out in Ref. \cite{Chen:2019ewa}. However, we consider only the simplest form of
	the K\"ahler potential.}.
Therefore, 
the canonical form of the kinetic terms  is obtained by 
changing the normalization of parameters \cite{Kobayashi:2018scp}.

For $\Gamma_3\simeq A_4$, the dimension of the linear space 
${\cal M}_k(\Gamma{(3)})$ 
of modular forms of weight $k$ is $k+1$ \cite{Gunning:1962,Schoeneberg:1974,Koblitz:1984}, i.e., there are three linearly 
independent modular forms of the lowest non-trivial weight $2$.
These forms have been explicitly obtained \cite{Feruglio:2017spp} in terms of
the Dedekind eta-function $\eta(\tau)$: 
\begin{equation}
\eta(\tau) = q^{1/24} \prod_{n =1}^\infty (1-q^n)~, 
 \quad\qquad  q= \exp \ (i 2 \pi  \tau )~,
 \label{etafunc}
\end{equation}
%
where $\eta(\tau)$ is a  so called  modular form of weight~$1/2$. 
In what follows we will use the following basis of the 
$A_4$ generators  $S$ and $T$ in the triplet representation:
\begin{align}
\begin{aligned}
S=\frac{1}{3}
\begin{pmatrix}
-1 & 2 & 2 \\
2 &-1 & 2 \\
2 & 2 &-1
\end{pmatrix},
\end{aligned}
\qquad \qquad
\begin{aligned}
T=
\begin{pmatrix}
1 & 0& 0 \\
0 &\omega& 0 \\
0 & 0 & \omega^2
\end{pmatrix}, 
\end{aligned}
\label{STbase}
\end{align}
%
where $\omega=\exp (i\frac{2}{3}\pi)$ .
The  modular forms of weight 2 transforming
as a triplet of $A_4$ can be written in terms of 
$\eta(\tau)$ and its derivative \cite{Feruglio:2017spp}:
\begin{eqnarray} 
\label{eq:Y-A4}
Y_1(\tau) &=& \frac{i}{2\pi}\left( \frac{\eta'(\tau/3)}{\eta(\tau/3)}  +\frac{\eta'((\tau +1)/3)}{\eta((\tau+1)/3)}  
+\frac{\eta'((\tau +2)/3)}{\eta((\tau+2)/3)} - \frac{27\eta'(3\tau)}{\eta(3\tau)}  \right), \nonumber \\
Y_2(\tau) &=& \frac{-i}{\pi}\left( \frac{\eta'(\tau/3)}{\eta(\tau/3)}  +\omega^2\frac{\eta'((\tau +1)/3)}{\eta((\tau+1)/3)}  
+\omega \frac{\eta'((\tau +2)/3)}{\eta((\tau+2)/3)}  \right) , \label{Yi} \\ 
Y_3(\tau) &=& \frac{-i}{\pi}\left( \frac{\eta'(\tau/3)}{\eta(\tau/3)}  +\omega\frac{\eta'((\tau +1)/3)}{\eta((\tau+1)/3)}  
+\omega^2 \frac{\eta'((\tau +2)/3)}{\eta((\tau+2)/3)}  \right)\,.
\nonumber
\end{eqnarray}
%
The overall coefficient in Eq.\,(\ref{Yi}) is 
one possible choice.
It cannot be uniquely determined.
The triplet modular forms of weight 2
have the following  $q$-expansions:
\begin{align}
{\bf Y^{(2)}_3}
=\begin{pmatrix}Y_1(\tau)\\Y_2(\tau)\\Y_3(\tau)\end{pmatrix}=
\begin{pmatrix}
1+12q+36q^2+12q^3+\dots \\
-6q^{1/3}(1+7q+8q^2+\dots) \\
-18q^{2/3}(1+2q+5q^2+\dots)\end{pmatrix}.
\label{Y(2)}
\end{align}
%
They satisfy also the constraint \cite{Feruglio:2017spp}:
\begin{align}
(Y_2(\tau))^2+2Y_1(\tau) Y_3(\tau)=0~.
\label{condition}
\end{align}

The  modular forms of the  higher weight, $k$, can be obtained
by the $A_4$ tensor products of  the modular forms  with weight 2,
 ${\bf Y^{(2)}_3}$, 
  as given in Appendix A.
  For weight 4, that is $k=4$, there are  five modular forms
   by the tensor product of  $\bf 3\otimes 3$ as:
\begin{align}
&\begin{aligned}
{\bf Y^{(4)}_1}=Y_1^2+2 Y_2 Y_3 \ , \quad
{\bf Y^{(4)}_{1'}}=Y_3^2+2 Y_1 Y_2 \ , \quad
{\bf Y^{(4)}_{1''}}=Y_2^2+2 Y_1 Y_3=0 \ , \quad
\end{aligned}\nonumber \\
\nonumber \\
&\begin{aligned} {\bf Y^{(4)}_{3}}=
\begin{pmatrix}
Y_1^{(4)}  \\
Y_2^{(4)} \\
Y_3^{(4)}
\end{pmatrix}
=
\begin{pmatrix}
Y_1^2-Y_2 Y_3  \\
Y_3^2 -Y_1 Y_2 \\
Y_2^2-Y_1 Y_3
\end{pmatrix}\ , 
\end{aligned}
\label{weight4}
\end{align}
where ${\bf Y^{(4)}_{1''}}$ vanishes due to the constraint of
 Eq.\,(\ref{condition}).
 For weight $6$, there are  seven modular forms
by the tensor products of  $A_4$ as:
\begin{align}
&\begin{aligned}
{\bf Y^{(6)}_1}=Y_1^3+ Y_2^3+Y_3^3 -3Y_1 Y_2 Y_3  \ , 
\end{aligned} \nonumber \\
\nonumber \\
&\begin{aligned} {\bf Y^{(6)}_3}\equiv 
\begin{pmatrix}
Y_1^{(6)}  \\
Y_2^{(6)} \\
Y_3^{(6)}
\end{pmatrix}
=
\begin{pmatrix}
Y_1^3+2 Y_1 Y_2 Y_3   \\
Y_1^2 Y_2+2 Y_2^2 Y_3 \\
Y_1^2Y_3+2Y_3^2Y_2
\end{pmatrix}\ , \qquad
\end{aligned}
\begin{aligned} {\bf Y^{(6)}_{3'}}\equiv
\begin{pmatrix}
Y_1^{'(6)}  \\
Y_2^{'(6)} \\
Y_3^{'(6)}
\end{pmatrix}
=
\begin{pmatrix}
Y_3^3+2 Y_1 Y_2 Y_3   \\
Y_3^2 Y_1+2 Y_1^2 Y_2 \\
Y_3^2Y_2+2Y_2^2Y_1
\end{pmatrix}\ . 
\end{aligned}
\label{weight6}
\end{align}
 By using these modular forms of weights $2, 4$ and $6$,
  we discuss  quark and lepton mass matrices.


\section{Quark mass matrices in the $A_4$ modular invariance}

Let us consider a $A_4$ modular invariant flavor model for quarks.
 There are freedoms for the assignments of irreducible representations and modular weights to quarks and Higgs doublets.

 The simplest one is to assign  the triplet of the $A_4$ group 
    to  three left-handed quarks, but three different singlets 
    $\bf (1,1'',1')$ of  $A_4$ to
    the  three right-handed quarks,
    ($u^c, c^c, t^c$) and  ($d^c, s^c, b^c$), respectively,
    where   the sum of weights of the left-handed and the  right-handed quarks is $-2$. The Higgs fields are supposed to be $A_4$ singlets with weight $0$.
Then,  three independent couplings appear in the superpotential of 
the up-type  and down-type quark sectors, respectively,
as follows:
 \begin{align}
 w_u&=\alpha_u u^c H_u {\bf Y^{(2)}_3} Q+
 \beta_u c^c H_u {\bf Y^{(2)}_3}Q+
 \gamma_u t^c H_u {\bf Y^{(2)}_3}Q~,\label{upquark} \\
w_d&=\alpha_d d^c H_d {\bf Y^{(2)}_3}Q+
\beta_d s^c H_d {\bf Y^{(2)}_3}Q+
\gamma_d b^c H_d {\bf Y^{(2)}_3}Q~,
\label{downquark}
\end{align}
where $Q$ is the left-handed $A_4$ triplet quarks,
and $H_q$ is the Higgs doublet.
The parameters $\alpha_q$,  $\beta_q$,  $\gamma_q$
are constant coefficients.
Assign  the  $A_4$ triplet $Q$ as $((d_L,u_L), (s_L,c_L), (b_L,t_L))$.
By using the decomposition of the $A_4$ tensor product in Appendix A, 
the superpotentials in Eqs.\,(\ref{upquark}) and (\ref{downquark}) give 
the mass matrix of quarks, which is written in terms of
modular forms of weight 2 as:
\begin{align}
\begin{aligned}
M_q=v_q
\begin{pmatrix}
\alpha_q & 0 & 0 \\
0 &\beta_q & 0\\
0 & 0 &\gamma_q
\end{pmatrix}
\begin{pmatrix}
Y_1 & Y_3& Y_2\\
Y_2 & Y_1 &  Y_3 \\
Y_3&  Y_2&  Y_1
\end{pmatrix}_{RL},     \qquad (q=u, d)~,
\end{aligned}
\label{matrixSM}
\end{align}
where  $\tau$ of the modular forms $Y_i(\tau)$ is  omitted.
 The constant $v_q\,(q=u,d)$ is the VEV of the neutral component of the Higgs field $H_q$.
Parameters $\alpha_q$,  $\beta_q$,  $\gamma_q$ are taken to be real without loss of generality, and they can be adjusted to the  observed quark masses. The remained parameter is only the modulus $\tau$. 
The numerical study of the quark mass matrix in Eq.\,(\ref{matrixSM}) is
 rather easy. 
However, it is impossible to  reproduce
observed  hierarchical three CKM mixing angles
by fixing one complex parameter $\tau$.
 
\begin{table}[h]
	\centering
	\begin{tabular}{|c||c|c|c|c|c|} \hline
		&$Q$&$(d^c,\,s^c,\,b^c)$& $(u^c,\,c^c,\,t^c)$&$H_q$&
		$\bf Y_3^{(2)}, \ Y_3^{(6)}, \ Y_{3'}^{(6)}$\\  \hline\hline 
		\rule[14pt]{0pt}{0pt}
		$SU(2)$&$\bf 2$&$\bf 1$&$\bf 1$&$\bf 2$&$\bf 1$\\
		$A_4$&$\bf 3$& \bf (1,\ 1$''$,\ 1$'$)&\bf (1,\ 1$''$,\ 1$'$)&$\bf 1$
		&$\bf 3, \qquad   3 , \qquad  3'$\\
		$-k_I$&$ -2$&$(0,\ 0,\ 0)$& $(-4,\, -4,\, -4)$&0&$k=2,\quad k=6,\quad k=6$ \\ \hline
	\end{tabular}
	\caption{Assignments of representations and  weights
		$-k_I$ for MSSM fields and  modular forms.
	}
	\label{tb:weight6}
\end{table}
 In order to obtain  realistic quark mass matrices,
  we use  modular forms of weight $6$  in Eq.\,(\ref{weight6}).
   As a simple model, we take modular forms of weight $6$ only for the up-type quark mass matrix
  while the down-type quark one is still given in terms of modular forms
   of weight 2 such as Eq.\,(\ref{matrixSM})
 \footnote{
 We also take  modular forms of weight $2$ for the charged lepton mass matrix
  to give a minimal number of  parameters in the lepton sector.}  
   .
   Then, we have six independent couplings in the superpotential of 
the up-quark sector as:
 \begin{align}
w_u&=\alpha_u u^c H_u {\bf Y^{(6)}_3} Q+
\alpha'_u u^c H_u {\bf Y_{3'}^{(6)}} Q+
\beta_u c^c H_u {\bf Y^{(6)}_3}Q+
\beta'_u c^c H_u {\bf Y_{3'}^{(6)}} Q \nonumber\\
& +\gamma_u t^c H_u {\bf Y^{(6)}_3}Q + \gamma'_u t^c H_q {\bf Y_{3'}^{(6)}} Q \,,
\end{align}
where assignments of representations and  weights
    for MSSM fields are given in  Table \ref{tb:weight6}.
    The up-type quark mass matrix is written as:
\begin{align}
\begin{aligned}
M_u=v_u
\begin{pmatrix}
\alpha_u & 0 & 0 \\
0 &\beta_u & 0\\
0 & 0 &\gamma_u
\end{pmatrix} \left [
\begin{pmatrix}
Y_1^{(6)} & Y_3^{(6)}& Y_2^{(6)} \\
Y_2^{(6)} & Y_1^{(6)} &  Y_3^{(6)} \\
Y_3^{(6)} &  Y_2^{(6)}&  Y_1^{(6)}
\end{pmatrix}
+ 
\begin{pmatrix}
g_{u1} & 0 & 0 \\
0 &g_{u2} & 0\\
0 & 0 &g_{u3}
\end{pmatrix}
\begin{pmatrix}
Y_1^{'(6)} & Y_3^{'(6)}& Y_2^{'(6)} \\
Y_2^{'(6)} & Y_1^{'(6)} &  Y_3^{'(6)} \\
Y_3^{'(6)} &  Y_2^{'(6)}&  Y_1^{'(6)}
\end{pmatrix}
\right ]_{RL},
\end{aligned}
\label{matrix6}
\end{align}
where $g_{u1}=\alpha'_u/\alpha_u$, $g_{u2}=\beta'_u/\beta_u$
and $g_{u3}=\gamma'_u/\gamma_u$ are complex parameters while
 $\alpha_u$,  $\beta_u$ and  $\gamma_u$  are  real.
 On the other hand, the down-type quark mass matrix is given as:  
 \begin{align}
 &\begin{aligned}
 M_d= v_d
 \begin{pmatrix}
 \alpha_d & 0 & 0 \\
 0 &\beta_d & 0\\
 0 & 0 &\gamma_d
 \end{pmatrix}
 \begin{pmatrix}
 Y_1 & Y_3& Y_2\\
 Y_2 & Y_1 &  Y_3 \\
 Y_3 &  Y_2&  Y_1
 \end{pmatrix}_{RL}\,.   
 \end{aligned} 
 \label{down}
 \end{align}
We will fix the modulus $\tau$ phenomenologically
 by using quark mass matrices in Eqs.\,(\ref{matrix6}) and (\ref{down}).

\section{Fixing $\tau$ by observed CKM}

In order to obtain the left-handed flavor mixing,
we calculate $M_d^{\dagger} M_d$ and  $M_u^{\dagger} M_u$.
At first, we take a random point of  $\tau$ and  $g_{ui}$
which are scanned  in the complex plane 
by generating random numbers. The modulus $\tau$ is scanned
in the fundamental domain of the modular symmetry.
In practice,  the  scanned range of 
${\rm Im } [\tau]$  is $[\sqrt{3}/2,2]$, in which the lower-cut $\sqrt{3}/2$ is at the cusp of the fundamental domain, and  the upper-cut $2$ is enough large for estimating  $Y_i$.
On the other hand,   ${\rm Re } [\tau]$ is scanned in
the fundamental domain  $[-1/2, 1/2]$ of the modular group. 
Supposing $|g_{ui}|$ is of order one, we scan them in $ [0,\,2]$ while these phases are scanned in $[-\pi,\pi]$.
Then, parameters $\alpha_q$,  $\beta_q$,  $\gamma_q$ ($q=u,d$)
are given in terms of $\tau$ and $g_q$ after inputting six quark masses.

Finally, we calculate three CKM mixing angles and the CP violating phase in terms of the model parameters $\tau$ and $g_{ui}$.
We keep the parameter sets, in  which the value of each observable is reproduced
within  the three times of $1\sigma$ interval of error-bars.
We continue this procedure to obtain enough points for plotting allowed region.

We input quark masses in order to constrain model parameters.
Since the modulus $\tau$ obtains the expectation value
 by the breaking of the modular invariance at the high mass scale,
   the quark masses are put  at the GUT scale.
   The observed masses and CKM parameters run to the GUT scale
 by the renormalization group equations (RGEs).
In our work, we adopt numerical values  of Yukawa couplings of quarks
  at the GUT scale $2\times 10^{16}$ GeV with $\tan\beta=5$ 
  in the framework of the minimal SUSY breaking scenarios
\cite{Antusch:2013jca, Bjorkeroth:2015ora}:
\begin{align}
\begin{aligned}
&y_d=(4.81\pm 1.06) \times 10^{-6}, \quad y_s=(9.52\pm 1.03) \times 10^{-5}, \quad y_b=(6.95\pm 0.175) \times 10^{-3}, \\
\rule[15pt]{0pt}{1pt}
&y_u=(2.92\pm 1.81) \times 10^{-6}, \quad y_c=(1.43{\pm 0.100}) \times 10^{-3}, \quad y_t=0.534\pm 0.0341  ~~,
\end{aligned}\label{yukawa5}
\end{align}
which give quark masses as $m_q=y_q v_H$ with $v_H=174$ GeV.

We also use the following CKM mixing angles to focus on parameter regions consistent with the 
experimental data  at the GUT scale $2\times 10^{16}$ GeV, where $\tan\beta=5$
is taken  \cite{Antusch:2013jca, Bjorkeroth:2015ora}:
\begin{align}
\begin{aligned}
&\theta_{12}^{\rm CKM}=13.027^\circ\pm 0.0814^\circ ~ , \qquad
\theta_{23}^{\rm CKM}=2.054^\circ\pm 0.384^\circ ~ ,  \qquad
\theta_{13}^{\rm CKM}=0.1802^\circ\pm 0.0281^\circ~ .
\end{aligned}\label{CKM}
\end{align}
Here $\theta_{ij}^{\rm CKM}$ is given in the PDG notation
of the CKM matrix $V_{\rm CKM}$ \cite{Tanabashi:2018oca}.
The  CP violating phase is also given as:
\begin{equation}
\delta_{CP}=69.21^\circ\pm 6.19^\circ~ ,
\label{CKMphase}
\end{equation}
  in the PDG notation.
The errors  in Eqs.\,(\ref{yukawa5}),  (\ref{CKM}) and (\ref{CKMphase}) represent $1\sigma$ interval.
The CKM elements $V_{ij}$ at the GUT scale  $2\times 10^{16}$ GeV  are given
 by using these angles and the phase.


In our model, we have four complex parameters, $\tau$, $g_{u1}$,
$g_{u2}$ and $g_{u3}$ after inputting six quark masses. 
These eight real parameters  are scanned to reproduce
 the observed three CKM mixing angles
and the CP violating phase 
with  three times  $1 \sigma$ error interval in Eqs.\,(\ref{CKM}) and (\ref{CKMphase})
\footnote{We take the observed values of CKM with  three times  $1\sigma$ intervals, which are almost $3\sigma$ in this case.}
.

We have succeeded to reproduce completely four observed  CKM elements 
in the parameter ranges of Table 2. The modulus $\tau$ is close to $i$,
 which is the fixed point of the modular symmetry.

\begin{table}[h!]
	\centering
	\begin{tabular}{|c|c|c|c|c|c|c|c|} \hline 
		\rule[14pt]{0pt}{0pt}
		$|{\rm Re} [\tau]|$
		&${\rm Im} [\tau]$ 
		&$|g_{u1}|$
		&${\rm Arg}\,g_{u1}$ 
		&$|g_{u2}|$
		&${\rm Arg}\,g_{u2}$
		&$|g_{u3}|$
		&${\rm Arg}\, g_{u3}$
		\rule[14pt]{0pt}{0pt}	 \\   		
		\hline \hline 
		$[0,\,0.09]$ &$[0.99,\,1.09]$ & $[0.01,\,0.86]$
		&$[-\pi,\, \pi]$&$[0.14,\,1.29]$ &  $[-2.3,\,1.6]$ & $[0.02,\,0.07]$& $[-\pi,\,\pi]$
		\rule[14pt]{0pt}{0pt}	\\
		\hline
	\end{tabular}
	\caption{Parameter ranges consistent with the observed CKM mixing angles and  CP phase $\delta_{CP}$.}
	\label{parameters}
\end{table}
\begin{figure}[h!]
	\vspace{2 mm}
	\begin{minipage}[]{0.47\linewidth}
		\includegraphics[{width=\linewidth}]{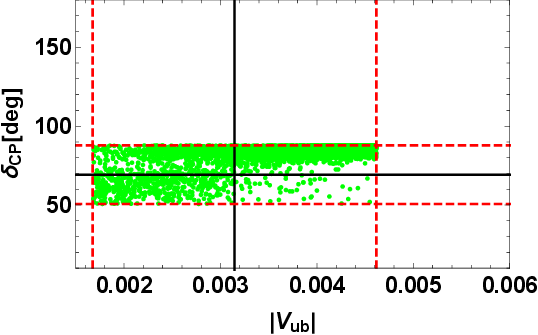}
		\caption{Distribution on  $|V_{ub}|$--$\delta_{CP}$ plane,
			where black lines denote  observed central values of 
			$|V_{ub}|$ and 	$\delta_{CP}$,   and  red dashed-lines denote
			three times    $1\sigma$ interval.	}
	\end{minipage}
	\hspace{5mm}
	\begin{minipage}[]{0.47\linewidth}
	\includegraphics[{width=\linewidth}]{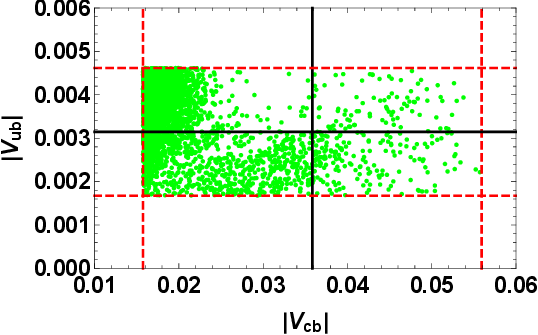}
	\caption{Distribution on $|V_{cb}|$--$|V_{ub}|$ plane
		where  black lines denote  observed central values of 
		$|V_{cb}|$ and 	$|V_{ub}|$,   and  red dashed-lines denote
		three times    $1\sigma$ interval.}
\end{minipage}
\end{figure}

In order to check the consistency of our quark mass matrices  and 
the  observed CKM,   
we show the calculated distribution 
on the $|V_{ub}|$--$\delta_{CP}$ plane at the GUT scale in Fig.\,1.
The calculated   $\delta_{CP}$ is uniformly distributed
  below the observed  central value of   $|V_{ub}|$
  while it is almost larger than the observed central value 
  for the upper-range of $|V_{ub}|$.


We also present the  distribution of  CKM elements $|V_{cb}|$ and $|V_{ub}|$ 
at the GUT scale  in Fig.\,2. 
 The magnitude of $|V_{ub}|$ is predicted 
to be in the whole region of the three times $1\sigma$ interval 
while  the calculated  $|V_{cb}|$ is mostly distributed
in the lower-range of the three times    $1\sigma$ interval.

\begin{wrapfigure}[18]{r}[0pt]{8.5 cm}
	\vspace{0mm}
		\includegraphics[{width=\linewidth}]{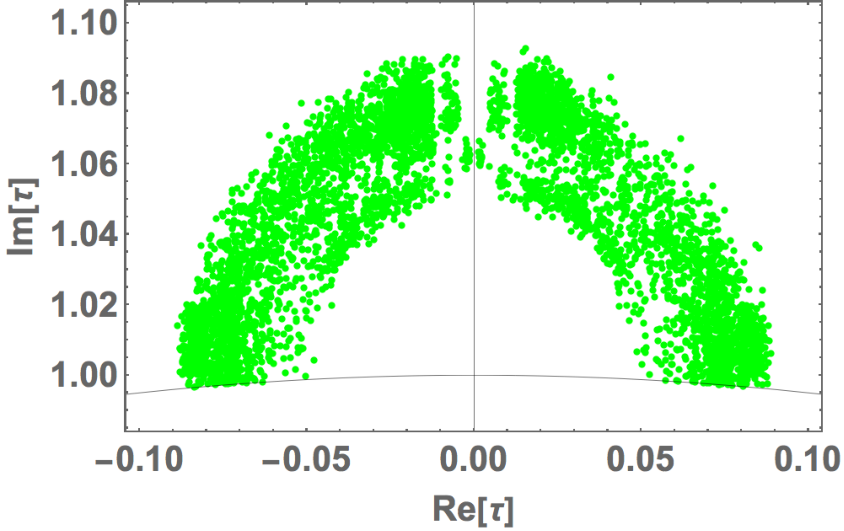}
	\caption{The allowed region  
	on ${\rm Re} [\tau]$--${\rm Im} [\tau]$ plane, where 
	observed  CKM mixing angles and  $\delta_{CP}$ are reproduced.
	The  solid curve is the boundary of the fundamental domain, $|\tau|=1$.}
\end{wrapfigure}


 In Fig.\,3,
 we show the plot  of ${\rm Re} [\tau]$ and ${\rm Im} [\tau]$,
 which will be compared   with the case of  leptons.
 The allowed region of the modulus $\tau$ is close to $i$,
 but is clearly  deviated from it.
 The modulus $\tau=i$ is the fixed point of the modular symmetry.
 Indeed, $\tau=i$ is invariant under the S transformation
 $\tau=-1/\tau$, where the subgroup $\mathbb{Z}_2^{S}=\{\rm I, S \}$ of $A_4$  
 is preserved.
  This region  of $\tau$ is discussed in connection with
    $\tau$ of the lepton mass matrices in the next secton.

 \begin{table}[h!]
 		\centering
 	\begin{tabular}{|c|c|} \hline 
 		\rule[14pt]{0pt}{0pt}	
 		$\tau$& $-0.038 + 1.05 \, i$  \\ 
 		\rule[14pt]{0pt}{0pt}
 		$g_{u1}$ &$-0.147 + 0.118\, i$ \\
 		\rule[14pt]{0pt}{0pt}
 		$g_{u2}$ & $ -0.091 - 0.425 \, i$ \\
 		\rule[14pt]{0pt}{0pt}
 		$g_{u3}$ & $0.027 + 0.0197\, i$  \\
 		\rule[14pt]{0pt}{0pt}
 	$\alpha_u/\gamma_u$ & $4.33\times 10^{-5}$  \\
 		\rule[14pt]{0pt}{0pt}
 	$\beta_u/\gamma_u$ & $ 3.85\times 10^{-3}$  \\
 		\rule[14pt]{0pt}{0pt}
 	$\alpha_d/\gamma_d$ & $1.45\times10^{-2}$  \\
 		\rule[14pt]{0pt}{0pt}
 		$\beta_d/\gamma_d$ & $4.26\times 10^{-3}$ \\
 		\rule[14pt]{0pt}{0pt}
 		$|V_{us}|$ & $0.225$ 	\\
 		\rule[14pt]{0pt}{0pt}
 		$|V_{cb}|$ & $  0.029$ 	\\
 		\rule[14pt]{0pt}{0pt}
 		$|V_{ub}|$ & $0.0030$ 	\\
 		\rule[14pt]{0pt}{0pt}
 		$\delta_{CP}$ & $76.9^\circ$ 	\\
 		\rule[14pt]{0pt}{0pt}
 		$\chi^2$ & $0.46$ 	\\
 		\hline
 	\end{tabular}
 	\caption{Numerical values of parameters and output of CKM parameters
 		at the best-fit point.	 
 	}
 	\label{sample}
 \end{table}
    In Table 3, 
    we  present  one parameter set and calculated   CKM elements,
    which is the best-fit point, 
    that is, its $\chi^2$ is  minimum.
    The magnitudes of  $g_{qi}$ are at most of order ${\cal O}(0.5)$.
     Ratios of  $\alpha_q/\gamma_q$ and $\beta_q/\gamma_q$ $(q=u,d)$
     correspond to the observed  quark mass hierarchy.
		
We also present the mixing matrices of  up-type quarks and
 down-type quarks for the sample of Table 3.
They are given as:

 \begin{align}
\begin{aligned}
V_u&\approx
\begin{pmatrix}
0.987& -0.017 + 0.142 \, i& 0.056- 0.039 \, i\\ 
		0.029+ 0.148 \, i&0.960& -0.236 - 0.014 \, i\\ 
-0.048 - 0.004 \, i& 	0.241- 0.021 \, i & 0.969
\end{pmatrix} \ , \\
V_d&\approx
\begin{pmatrix}
0.992& 0.083- 0.062\, i&	0.057- 0.043 \, i\\ 
-0.065- 0.049 \, i& 	0.962& -0.259 + 0.003\, i\\ 
	-0.076 - 0.058 \, i& 	0.251+ 0.002 \, i& 0.963
\end{pmatrix} \ ,
\end{aligned}
\label{V-I}
\end{align}
where $V_{\rm CKM}=V_u^\dagger \, V_d$.
It is noted that these are presented in the diagonal base of the generator $S$,
 where  we can see  the hierarchical structure of mixing.
 The  diagonal base of $S$ is realized
 by the unitary transformation of  $S$ in Eq.\,(\ref{STbase}), $V_{S}\, S \, V_{S}^\dagger={\rm diag}[1,-1,-1]$, where
 \begin{align}
V_{S}\equiv
\begin{pmatrix} 
\frac{1}{\sqrt{3}} &\frac{1}{\sqrt{3}} & \frac{1}{\sqrt{3}} \\
 \frac{2}{\sqrt{6}} & -\frac{1}{\sqrt{6}} & -\frac{1}{\sqrt{6}} \\
0 &-\frac{1}{\sqrt{2}} &\ \frac{1}{\sqrt{2}}
\end{pmatrix}\, .
\label{VS2}
 \end{align}
 Then,  the mixing matrix $V_q$ in the original base of $S$  is transformed to $V_S V_q$  because the quark mass matrix is transformed as 
 $V_S M_q^\dagger M_q V_S^\dagger$\,.


 In conclusion, our quark mass matrices
with the $A_4$ modular symmetry 
   reproduce the observed CKM mixing matrix very well  at nearby  $\tau=i$.
This successful result encourages us to investigate the lepton flavors
in the same framework. We discuss the lepton sector
with the $A_4$ modular symmetry in  Section 5.

\section{Lepton mass matrix in the $A_4$ modular invariance}
\subsection{Lepton mass matrix}

  The modular $A_4$ invariance also gives the lepton mass matrix
  in terms of the modulus $\tau$ which is common to both quarks and leptons if 
   flavors of quarks and leptons are originated from a same two-dimensional compact space.
  We assign the $A_4$ representation and   weight  for leptons  
   in Table 4, where
  the left-handed lepton doublets compose a $A_4$ triplet
  and the right-handed charged leptons are $A_4$ singlets.
  The weights of the leptons are assigned like  the down-type quarks in Table 1.
  
  \begin{table}[h]
  	\centering
  	\begin{tabular}{|c||c|c|c|c|c|} \hline
  		&$L$&$(e^c,\mu^c,\tau^c)$&$H_u$&$H_d$&$\bf Y_r^{(2)}, 
  		\ \   Y_r^{(4)}$\\  \hline\hline 
  		\rule[14pt]{0pt}{0pt}
  		$SU(2)$&$\bf 2$&$\bf 1$&$\bf 2$&$\bf 2$&$\bf 1$\\
  		$A_4$&$\bf 3$& \bf (1,\ 1$''$,\ 1$'$)&$\bf 1$&$\bf 1$&$\bf 3, \ \{3, 1, 1'\}$\\
  		$-k_I$&$ -2$&$(0,\, 0,\, 0)$ &0&0& \hskip -0.7 cm $2, \qquad 4$ \\ \hline
  	\end{tabular}	
  \caption{ Assignments of representations and  weights
  	$-k_I$ for MSSM fields and  modular forms of weight $2$ and $4$.
  	}
  	\label{tb:lepton}
  \end{table}
    Then, the superpotential of the  charged lepton mass term is given in terms of
  modular forms  of weight $2$, $\bf Y^{(2)}_3$ 
   since  weights of  the left-handed leptons and
     the right-handed charged leptons are $-2$ and $0$, respectively.
     It is given as:
     \begin{align}
     w_E&=\alpha_e e^c H_d {\bf Y^{(2)}_3}L+
     \beta_e \mu^c H_d {\bf Y^{(2)}_3}L+
     \gamma_e \tau^c H_d {\bf Y^{(2)}_3}L~,
     \label{chargedlepton}
     \end{align}
     where $L$ is the left-handed $A_4$ triplet leptons.
     Taking $(e_L, \mu_L,\tau_L)$ in the flavor base
    the charged lepton mass matrix $M_E$  is simply written  as:    
  \begin{align}
   \begin{aligned}
M_E=v_d \begin{pmatrix}
\alpha_e & 0 & 0 \\
0 &\beta_e & 0\\
0 & 0 &\gamma_e
\end{pmatrix}
\begin{pmatrix}
Y_1 & Y_3 & Y_2 \\
Y_2 & Y_1 & Y_3 \\
Y_3 & Y_2 & Y_1
\end{pmatrix}_{RL} \ ,
  \end{aligned}
   \label{ME(2)}
  \end{align}
where coefficients $\alpha_e$, $\beta_e$ and $\gamma_e$ are real parameters.

Suppose neutrinos to be Majorana particles.
By using the Weinberg operator, the superpotential of the neutrino mass term, $w_\nu$ is  given as:
\begin{align}
w_\nu&=-\frac{1}{\Lambda}(H_u H_u LL{\bf Y_r^{(4)}})_{\bf 1}~,
\label{Weinberg}
\end{align}
where $\Lambda$ is a relevant cutoff scale.
Since the  left-handed lepton doublet has weight $-2$, the superpotential
is given in terms of  modular forms of weight $4$, ${\bf Y_3^{(4)}}$,
${\bf Y_1^{(4)}}$ and  ${\bf Y_{1'}^{(4)}}$.

By using the tensor products of $A_4$, we have
\begin{align}
w_\nu &=\frac{v_u^2}{\Lambda}
\left [ 
\begin{pmatrix}
2\nu_e\nu_e-\nu_\mu\nu_\tau-\nu_\tau\nu_\mu\\
2\nu_\tau\nu_\tau-\nu_e\nu_\mu-\nu_\mu\nu_\tau\\
2\nu_\mu\nu_\mu-\nu_\tau\nu_e-\nu_e\nu_\tau
\end{pmatrix} \otimes
{\bf Y_3^{(4)}}  \right . \nonumber \\
& \left .  + \ 
(\nu_e\nu_e+\nu_\mu\nu_\tau+\nu_\tau\nu_\mu)
\otimes g_{\nu 1}{\bf Y_1^{(4)}}
+
(\nu_e\nu_\tau+\nu_\mu\nu_\mu+\nu_\tau\nu_e)
\otimes g_{\nu 2}{\bf Y_{1'}^{(4)}}
\right ]  \nonumber \\
=&\frac{v_u^2}{\Lambda}
\left[(2\nu_e\nu_e-\nu_\mu\nu_\tau-\nu_\tau\nu_\mu)Y_1^{(4)}+
(2\nu_\tau\nu_\tau-\nu_e\nu_\mu-\nu_\mu\nu_e)Y_3^{(4)}
+(2\nu_\mu\nu_\mu-\nu_\tau\nu_e-\nu_e\nu_\tau)Y_2^{(4)}\right .
\nonumber \\
& \left .  + \ 
(\nu_e\nu_e+\nu_\mu\nu_\tau+\nu_\tau\nu_\mu)
g_{\nu 1}{\bf Y_1^{(4)}}
+
(\nu_e\nu_\tau+\nu_\mu\nu_\mu+\nu_\tau\nu_e)
g_{\nu 2}{\bf Y_{1'}^{(4)}}
\right ]   \ , 
\end{align}
where ${\bf Y_3^{(4)}}$, ${\bf Y_1^{(4)}}$ and ${\bf Y_{1'}^{(4)}}$
are given in Eq.\,(\ref{weight4}), and  $g_{\nu 1}$, $g_{\nu 2}$ are complex parameters.
The neutrino mass matrix is written as follows:
\begin{align}
M_\nu=\frac{v_u^2}{\Lambda} \left [
\begin{pmatrix}
2Y_1^{(4)} & -Y_3^{(4)} & -Y_2^{(4)}\\
-Y_3^{(4)} & 2Y_2^{(4)} & -Y_1^{(4)} \\
-Y_2^{(4)} & -Y_1^{(4)} & 2Y_3^{(4)}
\end{pmatrix}
+g_{\nu 1} {\bf Y_{1}^{(4)}  }
\begin{pmatrix}
1 & 0 &0\\ 0 & 0 & 1 \\ 0 & 1 & 0
\end{pmatrix}
+g_{\nu 2} {\bf Y_{1'}^{(4)} }
\begin{pmatrix}
0 & 0 &1\\ 0 & 1 & 0 \\ 1 & 0 & 0
\end{pmatrix}
\right ]_{LL} \ .
\label{neutrinomassmatrix}
\end{align}
Then, the model parameters are
    $\alpha_e$, $\beta_e$, $\gamma_e$,   $g_{\nu 1}$ and $g_{\nu 2}$.
 Parameters $\alpha_e$, $\beta_e$ and  $\gamma_e$ are adjusted 
 by the observed charged lepton masses.
 Therefore, the lepton mixing angles, the Dirac phase  and Majorana phases
 are given by  $g_{\nu 1}$ and $g_{\nu 2}$ in addition to the value of $\tau$.
  Since  $\tau$ is restricted  in the narrow range for
   the quark sector
  as seen in Fig.\,3,
   we can give some predictions in the lepton sector.
   Practically, we scan $\tau$ in 
   $|{\rm Re} [\tau]|\leq 0.5 $  and ${\rm Im} [\tau]\leq 2$ like in the 
   analysis of quark mass matrices.

\subsection{Numerical results of leptons}
 We input charged lepton masses in order to constrain the model parameters.
We take Yukawa couplings of charged leptons 
at the GUT scale $2\times 10^{16}$ GeV,  where $\tan\beta=5$ is taken
as well as  quark Yukawa couplings
\cite{Antusch:2013jca, Bjorkeroth:2015ora}:
\begin{eqnarray}
y_e=(1.97\pm 0.024) \times 10^{-6}, \quad 
y_\mu=(4.16\pm 0.050) \times 10^{-4}, \quad 
y_\tau=(7.07\pm 0.073) \times 10^{-3},
\end{eqnarray}
where lepton masses are  given by $m_\ell=y_\ell v_H$ with $v_H=174$ GeV.
We also input  the   lepton mixing angles and neutrino mass parameters
which are given by NuFit 4.1 in Table 5 \cite{Esteban:2018azc}.
We investigate  two possible cases of neutrino masses $m_i$, which are
the normal  hierarchy (NH), $m_3>m_2>m_1$, and the  inverted  hierarchy (IH),
$m_2>m_1>m_3$.
Neutrino masses and  
the Pontecorvo-Maki-Nakagawa-Sakata (PMNS) matrix $U_{\rm PMNS}$ \cite{Maki:1962mu,Pontecorvo:1967fh} 
 are obtained by diagonalizing 
$M_E^\dagger M_E$ and $M_\nu^* M_\nu$.
We also investigate the effective mass for the $0\nu\beta\beta$ decay,
$\langle m_{ee} \rangle$ (see Appendix B)
and 
the sum of three neutrino  masses  $\sum m_i$  since
it is constrained by the recent cosmological data,
which is  the upper-bound $\sum m_i\leq 120$\,meV obtained at the 95\% confidence level
\cite{Vagnozzi:2017ovm,Aghanim:2018eyx}.
\begin{table}[h!]
	\begin{center}
		\begin{tabular}{|c|c|c|}
			\hline 
			\rule[14pt]{0pt}{0pt}
			\  observable \ &  $3\,\sigma$ range for NH  & $3\,\sigma$ range for IH \\
			\hline 
			\rule[14pt]{0pt}{0pt}
			$\Delta m_{\rm atm}^2$& \ \   \ \ $(2.436$--$ 2.618) \times 10^{-3}{\rm eV}^2$ \ \ \ \
			&\ \ $- (2.419$--$2.601) \times 10^{-3}{\rm eV}^2$ \ \  \\
			\hline 
			\rule[14pt]{0pt}{0pt}
			$\Delta m_{\rm sol }^2$& $(6.79$--$ 8.01) \times 10^{-5}{\rm eV}^2$
			& $(6.79$--$ 8.01)  \times 10^{-5}{\rm eV}^2$ \\
			\hline 
			\rule[14pt]{0pt}{0pt}
			$\sin^2\theta_{23}$&  $0.433$--$ 0.609$ & $0.436$--$ 0.610$ \\
			\hline 
			\rule[14pt]{0pt}{0pt}
			$\sin^2\theta_{12}$& $0.275$--$ 0.350$ & $0.275$--$ 0.350$ \\
			\hline 
			\rule[14pt]{0pt}{0pt}
			$\sin^2\theta_{13}$&$0.02044$--$ 0.02435$ & $0.02064$--$0.02457$ \\
			\hline 
		\end{tabular}
		\caption{The $3\,\sigma$ ranges of neutrino  parameters from NuFIT 4.1
			for NH and IH 
			\cite{Esteban:2018azc}. 
		}
		\label{DataNufit}
	\end{center}
\end{table}

  Let us  discuss numerical results for  NH of  neutrino masses.
 Since $\alpha_e/ \gamma_e$ and $\beta_e/ \gamma_e$ 
  are obtained by  the observed  charged lepton masses
  when $\tau$ is fixed, we input
    charged lepton masses to reduce free parameters.
  Parameters  $g_{\nu 1}$ and $g_{\nu 2}$   are constrained by
  four observed quantities;
  three mixing angles of leptons
  and observed mass ratio $\Delta m_{\rm sol}^2/\Delta m_{\rm atm}^2$.
  In practice, the  scanned ranges of 
   ${\rm Im } [\tau]$  and ${\rm Re } [\tau]$ are
    $[\sqrt{3}/2,2]$ and $[-1/2, 1/2]$, respectively, like for the quark sector. Neutrino couplings
    $|g_{\nu i}|$ are scanned in $ [0,\,10]$ while these phases are  in $[-\pi,\pi]$. Indeed, we have obtained $|g_{\nu 1}|=0.03$--$1.15$ and  
    $|g_{\nu 1}|= 0.63$--$1.22$ in our numerical calculations.
    
 \begin{wrapfigure}[22]{r}[0pt]{9 cm}
 	\vspace{0mm}
 	\includegraphics[{width=\linewidth}]{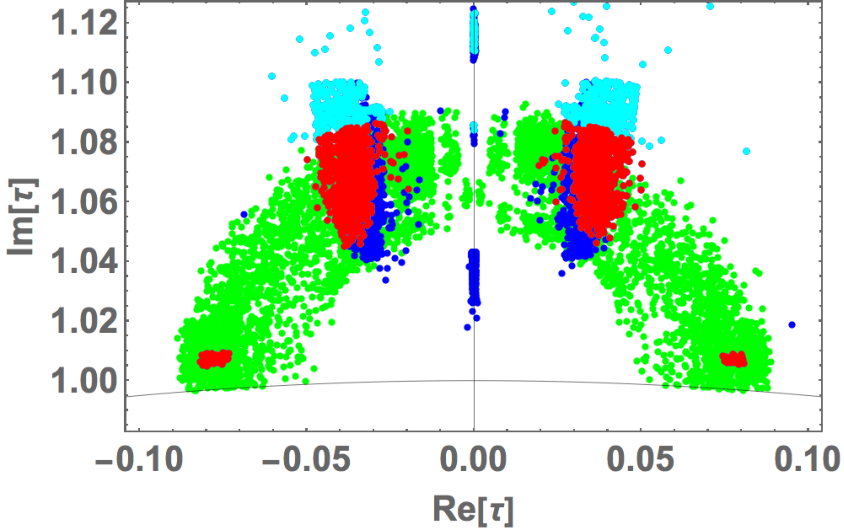}
 	\caption{Allowed regions of $\tau$.
 		PMNS mixing angles are reproduced
 		at cyan, blue and red points while  CKM  are reproduced at green  points. 
 		At cyan  (blue) points, the sum of neutrino masses 
 		is below (above) $120$\,meV. 
 		At red points, both CKM  and PMNS 	are reproduced 
 		with the sum of neutrino masses below $120$\,meV.
 		The  solid curve is the boundary of the fundamental domain, $|\tau|=1$.
 	}
 \end{wrapfigure}
   At first,  we show the allowed region  on the 
  ${\rm Re} [\tau]$--${\rm Im} [\tau]$ plane in Fig.\,4. Observed three mixing angles of leptons are reproduced at cyan, blue  and red points.
   The sum of neutrino masses 
  is consistent with  the cosmological bound $120$\,meV at cyan points,
  but not at blue points.
  At red points, both CKM  and PMNS 	are reproduced 
  with the sum of neutrino masses below $120$\,meV.
   For comparison, we add green points for quark CKM of Fig.\,3.
  Allowed points of leptons are almost in ${\rm Im} [\tau]\leq 1.12$
  and $|{\rm Re} [\tau]|\leq 0.1$.
 
 \begin{figure}[b!]
 	\begin{minipage}[]{0.47\linewidth}
 		\vspace{5mm}
 		\includegraphics[{width=\linewidth}]{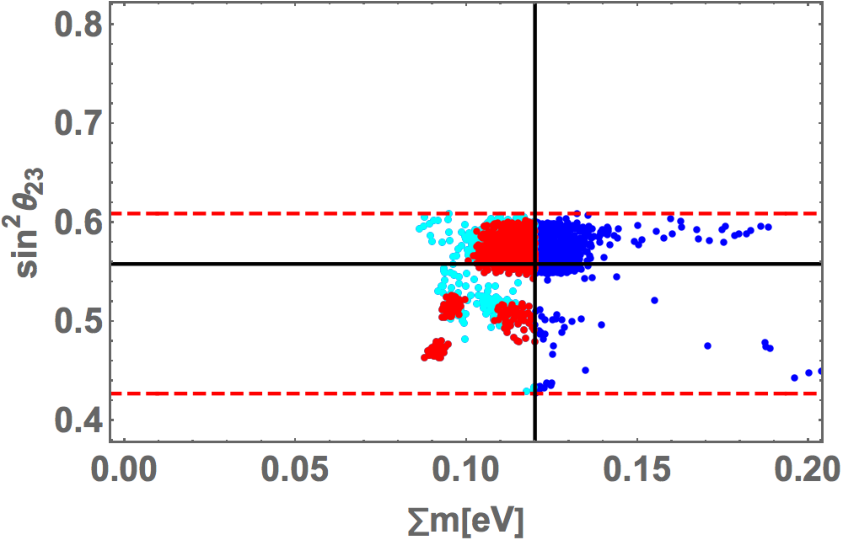}
 		\caption{Allowed region on   $\sum m_i$--$\sin^2\theta_{23}$ plane,
 			where the horizontal solid line denotes observed best-fit one,  red dashed-lines denote
 			the bound  of  $3\sigma$ interval,
 			and the vertical  line is the cosmological bound,	for NH.
 			Colors of points correspond to   $\tau$ in Fig.\,4.}
 	\end{minipage}
 	\hspace{5mm}
 	\begin{minipage}[]{0.47\linewidth}
 		\vspace{-5mm}
 		\includegraphics[{width=\linewidth}]{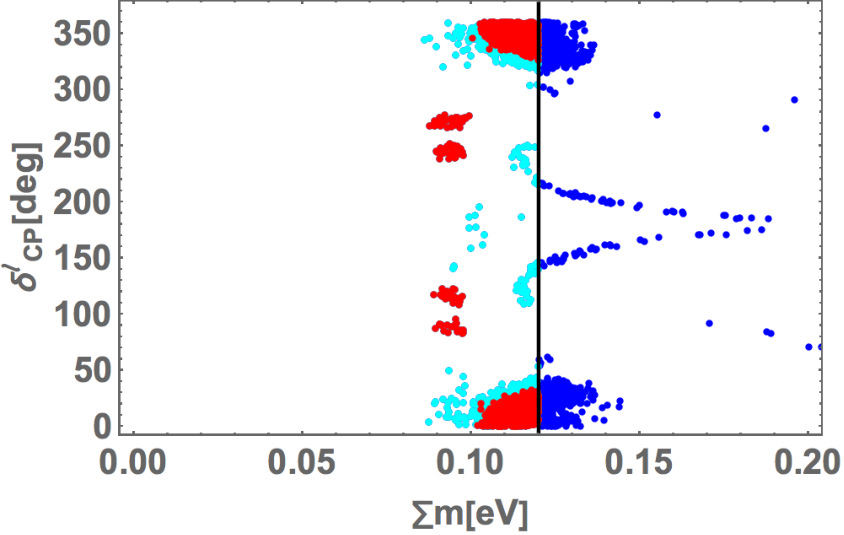}
 		\caption{Allowed region on the  $\sum m_i$--$\delta_{CP}^\ell$ plane,
 			where the vertical solid line is the cosmological bound,
 			for  NH .
 			Colors of points correspond to points of  $\tau$ in Fig.\,4}
 	\end{minipage}
 \end{figure}
  The common $\tau$ causes the feedback  in the quark sector.
  However, the tendency 
  of $|V_{ub}|$, $|V_{cb}|$ and $\delta_{CP}$ 
  are not so changed compared with Figs.\,1 and 2 in the common region of $\tau$.

We show the allowed region on  the $\sum m_i$--$\sin^2\theta_{23}$ plane
in Fig.\,5,
where colors (cyan, blue and red) of points correspond to points of  $\tau$ in Fig.\,4.
Our prediction of the sum of neutrino masses is  constrained by
the cosmological bound as seen in  Fig.\,5.
The minimal cosmological model, ${\rm \Lambda CDM}+\sum m_i$,
provides the upper-bound 
 $\sum m_i<120$\,meV
\cite{Vagnozzi:2017ovm,Aghanim:2018eyx} although it becomes  weaker when the data are analysed in the context of extended cosmological models \cite{Tanabashi:2018oca}.

The red region, that is the common $\tau$ region for quarks and leptons,
is constrained by the cosmological bound  $\sum m_i=120$\,meV.
 Then, the predicted  sum of neutrino masses  is  $87$--$120$\,meV.
The cyan region is inconsistent with  $\tau$ of quarks
while the blue one is excluded by  the  cosmological bound  $\sum m_i=120$\,meV,
although both are consistent with the data of NuFIT 4.1 \cite{Esteban:2018azc}.
The calculated  $\sin^2\theta_{23}$ of the red region  is distributed in restricted ranges.
Therefore, the precise measurement of $\sin^2\theta_{23}$ and  improving
the  bound of the sum of neutrino masses
provide crucial tests for our scheme.

\begin{figure}[t!]
	\begin{minipage}[]{0.47\linewidth}
		\includegraphics[{width=\linewidth}]{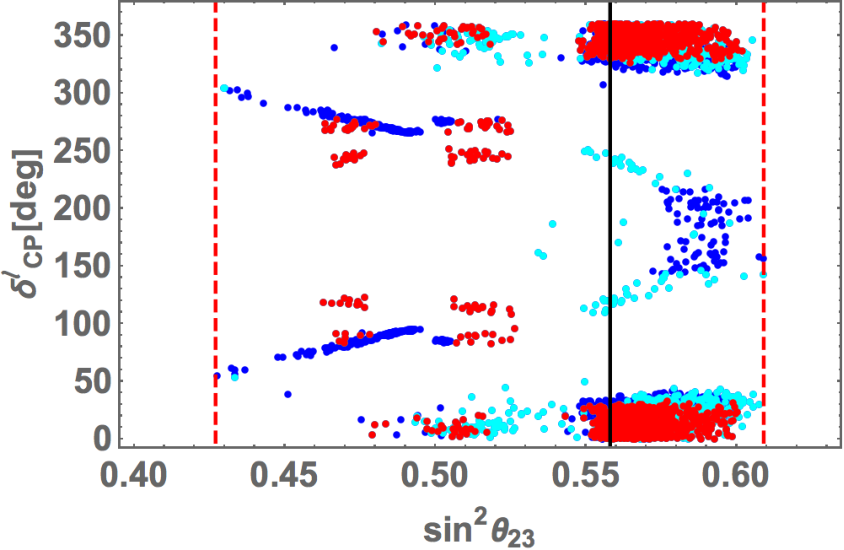}
		\caption{Predicted $\delta_{CP}^\ell$ versus  $\sin^2\theta_{23}$,
			where the black line denotes observed best-fit value of 
			$\sin^2\theta_{23}$,   and  red dashed-lines denote
			its upper(lower)-bound  of  $3\sigma$ interval
			for NH.
			Colors of points correspond to points of  $\tau$ in Fig.4.
		}
	\end{minipage}
	\hspace{5mm}
	\begin{minipage}[]{0.47\linewidth}
		\includegraphics[{width=\linewidth}]{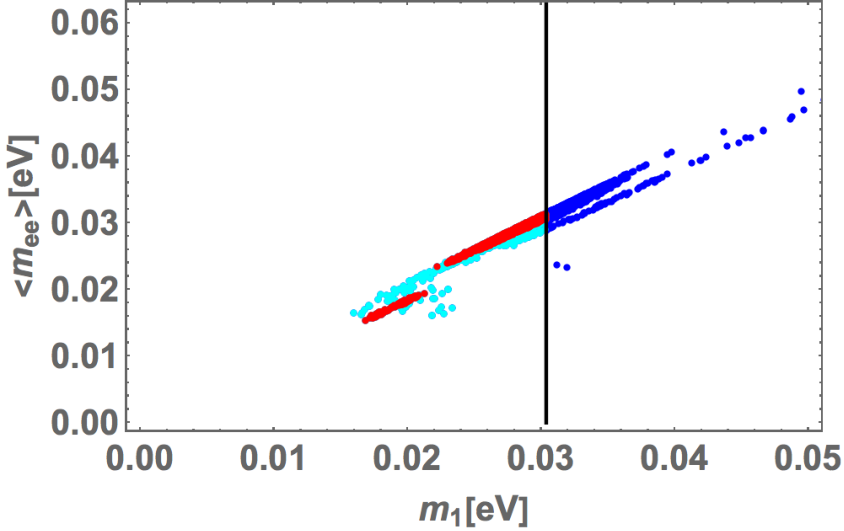}
		\caption{Predicted  effective mass $\langle m_{ee} \rangle$ of the $0\nu\beta\beta$ decay versus $m_1$ for  NH.
			The vertical  line is the upper-bound of $m_1$,
			which is derived from the cosmological bound, $120$\,meV.
			Colors of points correspond to points of  $\tau$ in Fig.4.}
	\end{minipage}
\end{figure}

We show the allowed region 
on the  $\sum m_i$--$\delta_{CP}^\ell$ plane in Fig.\,6.
In the region of red points, $\delta_{CP}^\ell$ is predicted to be in 
the restricted ranges,  
$0^\circ$--$50^\circ$, $80^\circ$--$100^\circ$, $110^\circ$--$130^\circ$,
$230^\circ$--$250^\circ$, $260^\circ$--$280^\circ$ and $310^\circ$--$360^\circ$.
If the cosmological bound for the sum of neutrino masses
 will be improved, for example, it is $100$\,meV,
$\delta_{CP}^\ell$ is predicted in  the distict range.


In Fig.\,7,  we  plot $\delta_{CP}^\ell$ versus $\sin^2\theta_{23}$  
in order to see their correlation. 
Since there is a  significant correlation between them, the precise measurement of $\sin^2\theta_{23}$ gives the clear prediction of  $\delta_{CP}^\ell$.

  
 We can also predict the effective mass 
 $\langle m_{ee}\rangle$ for the $0\nu\beta\beta$ decay versus 
 the lightest mass $m_1$ as seen in Fig.\,8.
 At the red point region, we predict
   $\langle m_{ee}\rangle=15$--$31$\,meV. 
  The predicted  $\langle m_{ee}\rangle$ larger than 
   $31$\,meV  is excluded by the sum of neutrino masses.
 
 \begin{table}[t!]
 	\centering
 	\begin{tabular}{|c|c|} \hline 
 			\rule[14pt]{0pt}{0pt}
 		$\tau$&   $-0.038 + 1.05 \, i$   \\ 
 		\rule[14pt]{0pt}{0pt}
 		$g_{\nu 1}$ &$ 0.061 + 0.274\, i$ \\
 		\rule[14pt]{0pt}{0pt}
 		$g_{\nu 2}$  &  $ 0.671 - 0.956\, i$ \\
 		\rule[14pt]{0pt}{0pt}
 		$\alpha_e/\gamma_e$ & $6.71\times 10^{-2}$  \\
 		\rule[14pt]{0pt}{0pt} 
 		$\beta_e/\gamma_e$ &  $1.24\times 10^{-3}$  \\
 		\rule[14pt]{0pt}{0pt}
 		$\sin^2\theta_{12}$ & $0.305$	\\
 		\rule[14pt]{0pt}{0pt}
 		$\sin^2\theta_{23}$ &  $ 0.565$	\\
 		\rule[14pt]{0pt}{0pt}
 		$\sin^2\theta_{13}$ &  $0.0216$	\\
 		\rule[14pt]{0pt}{0pt}
 		$\delta_{CP}^\ell$ &  $27.4^\circ$ 	\\
 		\rule[14pt]{0pt}{0pt}
 		$\sum m_i$ &  $118$\,meV 	\\
 		\rule[14pt]{0pt}{0pt}
 		$\langle m_{ee} \rangle$ &  $30.1$\,meV 	\\
 			\rule[14pt]{0pt}{0pt}
 		$\chi^2$ &  $0.64$ 	\\
 		\hline
 	\end{tabular}
 	\caption{Numerical values of parameters and output of PMNS parameters
 		at the best-fit point.}
 	\label{samplelepton}
 \end{table}

 In Table 6, 
 we  present  best-fit values of  parameters and outputs,
 where we input  the common value  $\tau=-0.038 + 1.05\, i$  in  Table 3.
 In our $\chi^2$ fit, $\delta_{CP}^\ell$ is not included,
 but it is only an output because  T2K and NO$\nu$A experiments
 have presented the best-fit values of $\delta_{CP}^\ell$ with opposite sign each other \cite{T2K:2020,Adamson:2017gxd}.
Our predicted $\delta_{CP}^\ell=27.4^\circ$ in Table 6 is rather small.
 The systematic $\chi^2$ fit of both quarks and leptons  will be needed
  by including  precise data of $\delta_{CP}^\ell$ since
    $\delta_{CP}^\ell\simeq \pm 90^\circ$ is also predicted in Fig.6.

 We also present the mixing matrices of   charged leptons and
 neutrinos for the best-fit sample of Table 6.
Those are given as:
 \begin{align}
 \begin{aligned}
 U_\ell&\approx
 \begin{pmatrix}
 0.989& 0.106- 0.073 \, i& 0.062- 0.043 \, i\\
 -0.087- 0.060 \, i&0.960 & -0.261 + 0.003\, i\\
 -0.087 - 0.060\, i&0.250+ 0.003\, i& 0.962
 \end{pmatrix} \ , \\
 U_\nu&\approx
 \begin{pmatrix}
 0.732& 0.465- 0.459 \, i&0.097- 0.166 \, i\\
 -0.333 - 0.267 \, i &0.228- 0.115 \, i & 0.868\\
 -0.415 - 0.332 \, i& 0.713& -0.448 - 0.094\,i
 \end{pmatrix} \ ,
 \end{aligned}
 \label{V-lepton}
 \end{align}
 where $U_{\rm PMNS}=U_\ell^\dagger \, U_\nu$.
 They are also given  in the diagonal base of the generator $S$
 in order to see  the hierarchical structure of mixing
  like in  the quark mixing matrix  in  Eq.\,(\ref{V-I}), 
 by using the unitary transformation of Eq.\,(\ref{VS2}). 
 It is noticed that the  mixing matrix of  charged leptons $U_\ell$
 is similar to the quark mixing matrices of  Eq.\,(\ref{V-I}).
 On the other hand, two large mixing angles appear
 in the neutrino mixing matrix $U_\nu$.


In our numerical calculations, we have not included  the RGE effects
 in the lepton mixing angles and neutrino mass ratio
 	$\Delta m_{\rm sol}^2/\Delta m_{\rm atm}^2$.
 We suppose that those corrections  are very small between 
 the electroweak  and GUT scales
  for NH of neutrino masses.
This assumption is  justified  well in the case of $\tan\beta\leq 5$
unless neutrino masses are almost degenerate \cite{Criado:2018thu}.


Finally, we  discuss briefly the case of IH of neutrino masses.
Indeed, there is a very small region of the common $\tau$ for  quarks and leptons,
which is marginal since the sum of neutrino masses is very close to 
the cosmological bound, $120$\,meV.
Therefore, we omit discussions of this case.

\newpage
\section{Summary}

We have studied both quark and lepton mass matrices
in the $A_4$ modular symmetry towards the unification of quark and lepton
flavors.
If  flavors of quarks and leptons are originated from a same two-dimensional compact space, 
 quarks and leptons have the same flavor symmetry and the same value of
the modulus  $\tau$.

For the quark sector, 
we have adopted modular forms of weights $2$ and $6$.
 We have presented the viable model for quark mass matrices,
 in which  the down-type quark mass matrix is constructed by modular forms of weight $2$ while  the up-type quark mass matrix is constructed by modular forms of weight $6$. 
 In the lepton sector,
 the charged lepton mass matrix is constructed  by modular forms of weight $2$
while modular forms of weight $4$ is used for the neutrino mass matrix,
  which is  generated by the Weinberg operator.
  
  The  viable region close to 
  $\tau=i$  is obtained in our quark mass matrices.
  Lepton mass matrices also work well  at nearby  $\tau=i$, which 
  overlaps with the one of the quark sector, for NH of neutrino masses.
   In the common $\tau$ region for quarks and leptons,
   the predicted  sum of neutrino masses is  $87$--$120$meV
    taking account of  its  cosmological bound.
   Since both the Dirac CP phase $\delta_{CP}^\ell$ and $\sin^2\theta_{23}$
   are correlated significantly with the sum of neutrino masses,
   improving its cosmological bound 
   provides  crucial tests for our scheme
   as well as  the precise measurement of $\sin^2\theta_{23}$ and 
    $\delta_{CP}^\ell$.
 The effective neutrino mass of the $0\nu\beta\beta$ decay 
is predicted to be  $\langle m_{ee}\rangle=15$--$31$meV.
 The IH  of  neutrino masses is almost excluded
by the cosmological bound of the sum of neutrino masses.

 It is remarked that  the common  $\tau$ is fixed at nearby $\tau=i$
in the fundamental domain of SL$(2,Z)$,
which suggests the residual symmetry $Z_2$ in the quark and lepton mass matrices.
Some corrections could violate the exact symmetry.
It is also emphasized that  
the spontaneous CP violation in Type IIB string theory  is possibly realized at nearby $\tau=i$, where the moduli
stabilization as well as the calculation of Yukawa couplings is performed in a controlled way \cite{Kobayashi:2020uaj}.
Thus, our phenomenological result of the modulus $\tau$ is favored 
in the theoretical investigation.
  

\section*{Acknowledgments}
This research was supported by an appointment to the JRG Program at the APCTP through the Science and Technology Promotion Fund and Lottery Fund of the Korean Government. This was also supported by
 the Korean Local Governments - Gyeongsangbuk-do Province and Pohang City (H.O.). H. O. is sincerely grateful for the KIAS member. 


\newpage
\appendix
\section*{Appendix}

\section{Tensor product of  $A_4$ group}
We take the generators of $A_4$ group for the triplet as follows:
\begin{align}
\begin{aligned}
S=\frac{1}{3}
\begin{pmatrix}
-1 & 2 & 2 \\
2 &-1 & 2 \\
2 & 2 &-1
\end{pmatrix},
\end{aligned}
\qquad 
\begin{aligned}
T=
\begin{pmatrix}
1 & 0& 0 \\
0 &\omega& 0 \\
0 & 0 & \omega^2
\end{pmatrix}, 
\end{aligned}
\end{align}
where $\omega=e^{i\frac{2}{3}\pi}$ for a triplet.
In this base,
the multiplication rule is
\begin{align}
\begin{pmatrix}
a_1\\
a_2\\
a_3
\end{pmatrix}_{\bf 3}
\otimes 
\begin{pmatrix}
b_1\\
b_2\\
b_3
\end{pmatrix}_{\bf 3}
&=\left (a_1b_1+a_2b_3+a_3b_2\right )_{\bf 1} 
\oplus \left (a_3b_3+a_1b_2+a_2b_1\right )_{{\bf 1}'} \nonumber \\
& \oplus \left (a_2b_2+a_1b_3+a_3b_1\right )_{{\bf 1}''} \nonumber \\
&\oplus \frac13
\begin{pmatrix}
2a_1b_1-a_2b_3-a_3b_2 \\
2a_3b_3-a_1b_2-a_2b_1 \\
2a_2b_2-a_1b_3-a_3b_1
\end{pmatrix}_{{\bf 3}}
\oplus \frac12
\begin{pmatrix}
a_2b_3-a_3b_2 \\
a_1b_2-a_2b_1 \\
a_3b_1-a_1b_3
\end{pmatrix}_{{\bf 3}\  } \ , \nonumber \\
\nonumber \\
{\bf 1} \otimes {\bf 1} = {\bf 1} \ , \qquad &
{\bf 1'} \otimes {\bf 1'} = {\bf 1''} \ , \qquad
{\bf 1''} \otimes {\bf 1''} = {\bf 1'} \ , \qquad
{\bf 1'} \otimes {\bf 1''} = {\bf 1} \  ,
\end{align}
where
\begin{align}
 T({\bf 1')}=\omega\,,\qquad T({\bf 1''})=\omega^2. 
\end{align}
More details are shown in the review~\cite{Ishimori:2010au,Ishimori:2012zz}.

\section{Majorana and Dirac phases and $\langle m_{ee}\rangle $
in  $0\nu\beta\beta$ decay }

Supposing neutrinos to be Majorana particles, 
the PMNS matrix $U_{\text{PMNS}}$~\cite{Maki:1962mu,Pontecorvo:1967fh} 
is parametrized in terms of the three mixing angles $\theta _{ij}$ $(i,j=1,2,3;~i<j)$,
one CP violating Dirac phase $\delta _\text{CP}$ and two Majorana phases 
$\alpha_{21}$, $\alpha_{31}$  as follows:
\begin{align}
U_\text{PMNS} =
\begin{pmatrix}
c_{12} c_{13} & s_{12} c_{13} & s_{13}e^{-i\delta^\ell_\text{CP}} \\
-s_{12} c_{23} - c_{12} s_{23} s_{13}e^{i\delta^\ell_\text{CP}} &
c_{12} c_{23} - s_{12} s_{23} s_{13}e^{i\delta^\ell_\text{CP}} & s_{23} c_{13} \\
s_{12} s_{23} - c_{12} c_{23} s_{13}e^{i\delta^\ell_\text{CP}} &
-c_{12} s_{23} - s_{12} c_{23} s_{13}e^{i\delta^\ell_\text{CP}} & c_{23} c_{13}
\end{pmatrix}
\begin{pmatrix}
1&0 &0 \\
0 & e^{i\frac{\alpha_{21}}{2}} & 0 \\
0 & 0 & e^{i\frac{\alpha_{31}}{2}}
\end{pmatrix},
\label{UPMNS}
\end{align}
where $c_{ij}$ and $s_{ij}$ denote $\cos\theta_{ij}$ and $\sin\theta_{ij}$, respectively.

The rephasing invariant CP violating measure of leptons \cite{Jarlskog:1985ht,Krastev:1988yu}
is defined by the PMNS matrix elements $U_{\alpha i}$. 
It is written in terms of the mixing angles and the CP violating phase as:
\begin{equation}
J_{CP}=\text{Im}\left [U_{e1}U_{\mu 2}U_{e2}^\ast U_{\mu 1}^\ast \right ]
=s_{23}c_{23}s_{12}c_{12}s_{13}c_{13}^2\sin \delta^\ell_\text{CP}~ ,
\label{Jcp}
\end{equation}
where $U_{\alpha i}$ denotes the each component of the PMNS matrix.

There are also other invariants $I_1$ and $I_2$ associated with Majorana phases
\begin{equation}
I_1=\text{Im}\left [U_{e1}^\ast U_{e2} \right ]
=c_{12}s_{12}c_{13}^2\sin \left (\frac{\alpha_{21}}{2}\right )~, \quad
I_2=\text{Im}\left [U_{e1}^\ast U_{e3} \right ]
=c_{12}s_{13}c_{13}\sin \left (\frac{\alpha_{31}}{2}-\delta^\ell_\text{CP}\right )~.
\label{Jcp}
\end{equation}
We can calculate $\delta^\ell_\text{CP}$, $\alpha_{21}$ and $\alpha_{31}$ with these relations by taking account of 
\begin{eqnarray}
&&\cos\delta^\ell_{CP}=\frac{|U_{\tau 1}|^2-
	s_{12}^2 s_{23}^2 -c_{12}^2c_{23}^2s_{13}^2}
{2 c_{12}s_{12}c_{23}s_{23}s_{13}}~ , \nonumber \\
&&\text{Re}\left [U_{e1}^\ast U_{e2} \right ]
=c_{12}s_{12}c_{13}^2\cos \left (\frac{\alpha_{21}}{2}\right )~, \qquad
\text{Re}\left [U_{e1}^\ast U_{e3} \right ]
=c_{12}s_{13}c_{13}\cos\left(\frac{\alpha_{31}}{2}-\delta^\ell_\text{CP}\right )~.
\end{eqnarray}
In terms of this parametrization, the effective mass for the $0\nu\beta\beta$ decay is given as follows:
\begin{align}
\langle m_{ee}	\rangle=\left| m_1 c_{12}^2 c_{13}^2+ m_2s_{12}^2 c_{13}^2 e^{i\alpha_{21}}+
 m_3 s_{13}^2 e^{i(\alpha_{31}-2\delta^\ell_{CP})}\right|  \ .
\end{align}


\end{document}